\documentstyle[aaspptwo]{article}
\lefthead{Strickman, {et al.}}
\righthead{The Vela Pulsar}
\begin{document}
\title{An RXTE Observation of the Vela Pulsar:  Filling in the X-ray Gap}

\author{M.S.~Strickman\altaffilmark{1}}
\affil{Naval Research Laboratory}
\authoraddr{Washington, DC 20375-5352} 
\author{A.K.~Harding} 
\affil{NASA Goddard Space Flight Center}
\authoraddr{Greenbelt, MD 20771} 
\and 
\author{O.C.~deJager}
\affil{Potchefstroom University}
\authoraddr{Potchefstroom 2520, South Africa}
\altaffiltext{1}{E-mail: strickman@osse.nrl.navy.mil}
\begin{abstract} 
We have detected pulsed emission from the Vela pulsar at 2 - 30 keV 
during a 93 ks observation with the {\it Rossi X-Ray Timing Explorer} 
(RXTE).  The RXTE pulse profile shows two peaks, that are roughly in 
phase with the EGRET peaks, but does not show any significant interpeak 
emission.
The phase of Peak 2 is energy dependent, moving to higher phase with 
increasing energy in the RXTE band, in phase alignment with the second
optical pulse in the lowest energy band of 2 - 8 keV and in phase
alignment with the 
second EGRET pulse in the highest energy band of 16-30 keV.
The average pulse spectrum joins smoothly to the high energy spectrum
of OSSE, COMPTEL and EGRET, although the spectrum of Peak 1 is significantly
harder than that of Peak 2.  A break or turnover in the spectrum 
around 50 keV, previously suggested by OSSE data, is now clearly defined.  
The RXTE spectrum falls several orders of
magnitude below the ROSAT emission in the 0.1 - 2 keV band, suggesting 
a thermal origin for the ROSAT pulse.  

\end{abstract}


\section{Introduction}

The Vela Pulsar (PSR B0833-45) is representative of a group of pulsars, with
ages in the range $\sim10^{4} - 10^{5}$ years, that are surrounded by 
synchrotron
nebulae.  Of these, the Vela Pulsar, PSR B1706-44, PSR B1951+32 and
possibly PSR B0656+14 are also
high energy $\gamma$-ray emitters (Thompson et al. \markcite{thomp97} 1997).
In addition to these four pulsars, a number of other ``Vela-type'' pulsars
have been observed to be steady point sources of soft X-rays (Becker \& 
Tr\"{u}mper \&  \markcite{becke97} 1997).  Because of its proximity and low spin-down
age within this group, the Vela Pulsar is the easiest to observe across the
high energy spectrum, being the brightest object in the sky
above 100 MeV.  Vela, as well as PSR B0656+14, exhibit soft 
(less than 1.2 keV) X-ray 
pulsations, detected by ROSAT (\"{O}gelman \markcite{ogelm93} 1993). 
Recent observations of Vela by the OSSE (Strickman et al. \markcite{stric96}
1996) and COMPTEL (Bennett et al. \markcite{benne94} 1994) experiments on
board Compton Gamma Ray Observatory have filled in part of the gap between
soft X-rays and high energy $\gamma$-rays, extending the pulsed spectrum down
to 50 keV.  However, the 2-50 keV hard X-ray band has remained without a
detection of pulsed emission.  

Characterizing the pulsed emission in the 2-50 keV range is important for
several reasons.  The pulsed X-ray flux detected by ROSAT was extremely soft
and well described by a black body model.  Since the pulsed spectra above
50 keV are clearly nonthermal, measurements in the intervening energy band are
required in order to determine the nature of the transition from a dominant thermal to nonthermal emission mechanism.  In addition, the pulse profile 
above 50 keV is similar to that observed above 100 MeV, while the pulse profile in the black body region below 1.2 keV is significantly different.  
Determination of the pulsar profile in the 2 - 50 keV band is clearly
also of interest.  Finally, both polar cap (e.g. Daugherty \& Harding \markcite{daugh96} 1996, Harding \& Daugherty\markcite{hard99} 1999) 
and outer gap (e.g. Romani
\& Yadigaroglu \markcite{roman95} 1995, Romani \markcite{roman96} 1996) models
make distinct predictions concerning spectra and pulse profiles in this energy
region.  Hence, 2 - 50 keV observations there may constrain one or both model 
catagories.

In order to address this hole in spectral coverage, we have performed a
sensitive Vela pulsar observation using the Proportional Counter Array (PCA)
experiment on board the Rossi X-ray Timing Explorer (RXTE; see Bradt,
Rothschild \& Swank \markcite{bradt93} 1993 for a description).  In this
paper, we report on the results of this observation, and discuss some of
their implications for the questions mentioned above.  Detailed comparisons
with model predictions will be presented in a future paper (but see
Harding \& Daugherty\markcite{hard99} 1999).

\section{Observations and Analysis} \label{sec:obs}

We observed the Vela Pulsar with RXTE during the intervals indicated in
Table~\ref{tbl:velaobs}, with a total good exposure of 93 ksec.  PCA data were
collected in GoodXenon event-by-event mode and hence were available with full
time and energy resolution.  For a pulsar
ephemeris, we used the Princeton Pulsar Database (Arzoumanian et al.
\markcite{arzou92} 1992), which is available  via the WWW at
http://pulsar. princeton.edu/ftp/gro/psrtime.dat.  For our analysis, we used
the entry valid from MJD 50402 -- 50479.  Table~\ref{tbl:velaephem} lists the
frequency and derivatives from the database, in addition to ``${\rm t_{0
geo}}$,'' the absolute time epoch.  The latter includes the Vela Pulsar
correction of 13.4 msec, which accounts for variations in dispersion measure
and in the definition of ``pulse peak.''  See the database web site for a
detailed discussion of ``${\rm t_{0 geo}}$.''

Epoch folding analysis was accomplished using pre-release versions of the
FASEBIN family of tools included in the FTOOLS version 4 release.  These tools
perform the actual phase coherent epoch fold (FASEBIN), add individual days
together (FBADD), subtract ``off-pulse'' regions (FBSUB), and create light
curves (FBSSUM) and phase-resolved spectra (FBFSUM).  Subsequent analyses were
performed using XSPEC and IDL.  We have examined the time averaged light curves and phase
resolved spectra, the variations in the time averaged light curve features
with energy, and have looked for temporal variability in both flux and
spectral shape.

In order to study the time averaged light curve as a function of energy, we
folded the event stream into 88-bin (approximately 1 msec/bin) phase
histograms which represent average light curves for each of
three broad energy bands:  2 -- 8 keV, 8 -- 16 keV and
16 -- 30 keV.  The results are shown in Figure~\ref{fig:pcalcs}, where the
origin of the phase axis represents the centroid of the radio peak.  In the
lowest band, we used layer 1 of the proportional counter only, while in the
other bands we used all three xenon layers summed together.  The light curves
have been divided into five phase regions, as listed in
Table~\ref{tbl:phsregions}. The regions, entitled ``Peak 1 Precursor'', ``Peak
1'',
``Peak 2'', ``Off Pulse 1'' and ``Off Pulse 2'' and abbreviated on the figure, were chosen
by eye to assure that changes in peak shape with energy would not cause the
``off pulse'' regions to become contaminated with peak flux at any energy.
In all cases, ``peak 1'' and ``peak 2'' refer to the order of
the peaks as they follow the radio peak.  We have examined smaller regions but
have found that the statistical precision of the data do not support smaller
regions than the ones used here. Regions OP1 and OP2, the two regions
identified as ``off-pulse,'' were chosen as such since they are consistent
with the light curve minimum, despite the fact that the EGRET result indicates
that at gamma-ray energies there is modulated flux in the OP1 phase region. 

To better characterize the behavior of light
curve features with energy, we modeled the light curve using a constant
background and two Lorentzian-shaped peaks.  As we were primarily interested
in peak width and separation, and due to limited statistics, the model did not
include the peak 1 precursor region, and that region was excluded from the
fits.  This model represents the 2 -- 8 keV and 8 -- 16 keV data quite well
(reduced $\chi^{2}$ of 1.11 and 1.16 for 58 degrees of freedom respectively). 
In the 16 -- 30 keV band, bin-to-bin fluctuations result in a reduced
$\chi^{2}$ of 1.44 for 58 dof.  However, the model is quite acceptable in this
band if we fold the data into 44 phase bins rather than 88.  The resulting fit
parameters are essentially identical to those from the original 88-bin best
fit model.

Phase resolved spectra were created using the straightforward ``on pulse''
minus ``off pulse'' technique, then time normalized to the average over the
entire light curve.  This normalization removes any bias that might be
introduced by having to make the peak 2 phase region wider than necessary at
any given energy.  Spectra were kept in native PCA channels, with layer 1 only
used for channels less than 8 keV and the sum of all 3 Xenon layers used above
8 keV.  Analyses of these spectra were performed using XSPEC
together with PCA
responses produced with the standard FTOOLS.  

Finally, we also performed analyses of the light curve and spectra on a
day-by-day basis.  These were performed similarly to the time-integrated
analyses discussed above, except that poorer statistics limited us to
relatively crude studies of broad band pulsed flux and spectrum shape
as a function of time.

\section{Results} \label{sec:anal}

We detect pulsed emission in all three of the broad energy bands shown in 
Figure~\ref{fig:pcalcs}.  Using the Maximum Likelihood Ratio (MLR) test of 
Li \& Ma \markcite{lima83} (1983), we have assessed the probability (expressed
as a number of Gaussian sigmas) that each of three light
curve features in each band is consistent with random fluctuations in the
data.  As indicated in Table~\ref{tbl:velamlr}, both peaks are significantly
detected in all three bands.  The peak 1 precursor region is significant (and,
at that, only marginally), in the 2 -- 8 keV band only.  We find no significant 
pulsed emission above background in either the OP1 region (between the
peaks) or the OP2 region (outside the peaks).  The lack of pulsed emission
in the OP1 region is in constrast to the EGRET detection of significant
interpeak emission in the Vela pulsar profile above 30 MeV.

\begin{figure}
\plotone{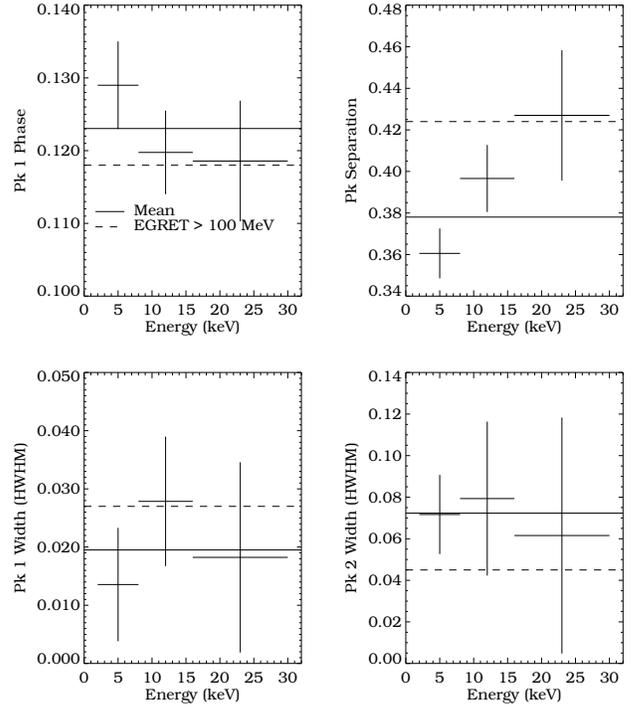}
\caption{Light curve peak properties vs. energy.  Parameters are determined by
fitting a 2 Lorentzian peak model to the lightcurves for each PCA energy band.
The solid line is the mean of the PCA results averaged over energy.  The
dashed line is the value of the same peak parameter above 100 MeV from EGRET (Kanbach et al. 
1994).}   
\label{fig:pkfits}
\end{figure}

In order to examine the behavior of the light curve features with energy and
to quantitatively compare them with similar features found in high energy
gamma-rays, we fit Lorentzian-shaped peak models, discussed in the previous
section, to the observed light curves in all three bands. 
Figure~\ref{fig:pkfits} shows the results of these fits.  In the cases of
peak~1 absolute phase and the widths of both peak 1 and peak 2, we see no
evidence
for variation of the parameter with energy.  Additionally, the fitted
parameter values are consistent with the same parameter observed by EGRET
above 100 MeV (Kanbach et al. \markcite{kanba94} 1994), as shown by the dashed
lines in the figure.

However, we do see evidence that the peak separation differs from that
observed by EGRET, especially at low energies.  In
particular, the peaks appear to be significantly closer together at 2 -- 8 keV
than observed by EGRET.  Above 8 keV, peak separations are consistent with
those observed by EGRET.  The reduced $\chi^{2}$ for modeling the peak
separation vs. energy by the mean PCA result averaged over all three bands is 
2.9, which corresponds to a 5\% probability of the result being produced
randomly.  Hence we cannot say conclusively that the separation varies from
PCA data alone.  However, comparing the PCA results to the EGRET peak
separation yields a reduced $\chi^{2}$ of 15.4 corresponding to a random
probability of $2\times 10^{-7}$, which is clearly significant.  We therefore
conclude from the combination of PCA and EGRET results that the peak
separation increases with increasing energy in the X-ray regime, although
there is no statistically significant evidence for change above 8 keV.

Peak minus Off-Pulse spectra were created for the peak 1 precursor, peak 1 and
peak 2 phase regions.  Power-law with photoelectric absorption fits have been
performed for each spectrum.  Fits have used the XSPEC PEGPWRLW photon
spectrum model with normalization set to 10 keV and the XSPEC WABS absorption
model with a column density fixed at $1\times10^{20}$ atoms-cm$^{-2}$, based
on ROSAT results (\"{O}gelman \markcite{ogelm93}1993).  The resulting best fit
photon power-law indices $\alpha$ are: Peak~1 precursor, $\alpha = 1.57\pm0.29$, $\chi^{2}/dof = 1.35$; Peak~1, $\alpha = 0.68\pm0.14$, $\chi^{2}/dof = 1.19$; 
and Peak~2, $\alpha = 1.17\pm0.12$,
$\chi^{2}/dof= 1.49$.  While this simple model does not represent the spectra
perfectly (although some portion of $\chi^{2}$ stems from known systematics in
the PCA response), the resulting indices are representative of the relative
hardness of each spectrum.  Comparing these indices to their mean, we can
reject the hypothesis that they are drawn from the same distribution at about
the $3\sigma$ level.  The PCA peak~1 and peak~2 spectra (with channels summed
together for clarity of display) are shown in the context of optical and other
X-ray and gamma-ray observations in Figure~\ref{fig:velaspec}.  They clearly
continue the hardening trend from higher to lower energies first observed
by OSSE and COMPTEL.  

We have also produced a time history of the pulsed emission on a day-by-day
basis.  We see marginal evidence that the total pulsed flux may have
disappeared during the first observing day in the 2 -- 8 keV band and during
the sixth observing day in the 8 -- 30 keV band.  In the former case, the
probability of such a variation being a random fluctuation, as determined by
$\chi^{2}$, is $6\times10^{-4}$, while in the latter it is $2\times10^{-3}$. 
The probability of seeing two such fluctuations in 16 trials from randomly
distributed data is $10^{-5}$. Statistics do not allow further subdivision
into individual light curve components and/or shorter time intervals.  We have
examined a number of possible systematics that could have caused this
phenomenon.  In particular, because the off-pulse phase region is used as
background, normal PCA background concerns for weak sources do not apply.  We
refolded the data with an alternative ephemeris (based on some but not all of
the same radio data) and obtained the same result.  In addition, the fact that
pulsations are observed in some energy band on all days mitigates against
epoch folding problems.  Likewise, we find no evidence for deadtime
variations, which would effect all energies equally if they were present in
any event.  Due to limited statistics and continuing concerns over possible
systematic effects, we cannot draw any firm conclusions about variability of
pulsed emission from these results.  However, we have recently completed a
much deeper RXTE observation of the Vela Pulsar, and will carefully examine it for
time variability on daily and other time scales.  None of the existing pulsar
models have addressed the question of high-energy flux variability on day to
week timescales.

\section{Discussion} \label{sec:disc} 

\begin{figure}
\plotone{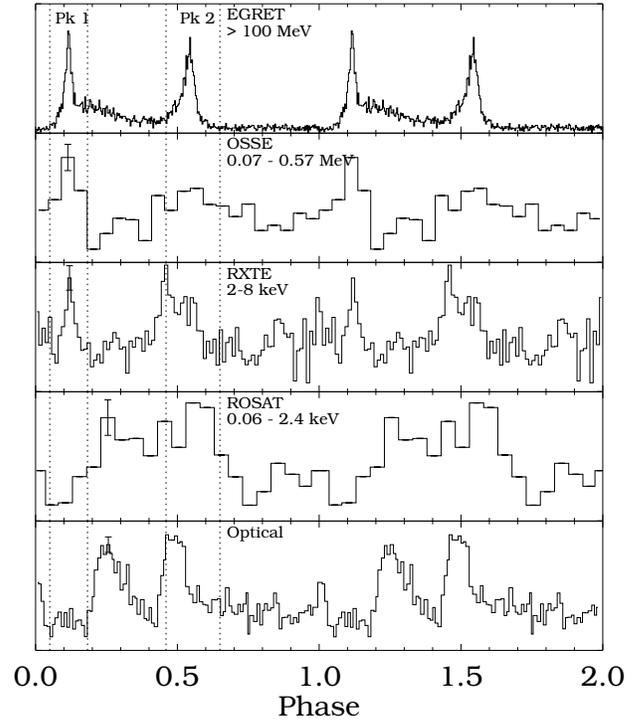}
\caption{Epoch-folded light curves of pulsed high-energy and optical
emission from the Vela Pulsar. 
In addition to the 2 - 8 keV PCA light curve from this work, data shown
include results from
EGRET, OSSE, ROSAT, and optical observations. 
See Figure~3 for references.  In all
cases, the origin of the phase axis is the phase of the centroid of the radio
pulse.}   
\label{fig:velalcs}
\end{figure}

Our detection of pulsed emission in the range 2 - 30 keV from the Vela
pulsar has succeeded in filling in the gap over the hard X-ray energy
band and has important implications for the origin of the high energy
emission and its connection to lower energy bands.  The pulse profile 
measured by RXTE from 2 - 30 keV shows two peaks separated by 0.4
in phase, similar to the profile measured at high energies. 
Figure~\ref{fig:velalcs}
shows the Vela pulse profiles at energies ranging from
optical to high-energy $\gamma$-rays, including the RXTE profile averaged
over the 2 - 8 keV energy band.  The phase of the RXTE Peak 1 is the 
same as that of EGRET and OSSE, within errors.  The Vela profile above
2 keV thus shows a very similar behavior to the Crab pulse profile,
but does not, however, duplicate the phase alignment of the Crab at lower 
energies.  

The spectra of Peaks 1 and 2 also join smoothly, 
within errors, to the higher energy spectrum of OSSE, COMPTEL and EGRET,
indicating that the RXTE emission is a continuation of the non-thermal
magnetospheric emission to lower energy.  The RXTE spectrum, however,
clearly indicates a break at around 50 keV, confirming a feature that
was suggested by the low-energy OSSE points.  Recent RXTE upper limits 
on pulsed emission from PSR B1706-44 (Ray et al. 1999) and from PSR B1951+32 
(Cheng \& Ho 1997) also require similar low-energy breaks or turnovers in the 
high-energy spectra of these pulsars.  Such a low-energy flattening
or turnover is predicted by the polar cap cascade model (Daugherty \&
Harding 1996; Harding \& Daugherty 1999).  In this model the 
synchrotron spectrum of
the pair cascade will exhibit a turnover at the local cyclotron frequency,
blueshifted by the relativistic motion of the pairs along the magnetic
field lines by a factor $\Gamma_{\pm} \sim 20$. The energy of the 
turnover is a strongly increasing function of local magnetic field strength
and thus provides a measure of the height of the emission above the
neutron star surface.  In the case of Vela, a turnover at 50 keV would
indicate that the polar cap cascade emission is occurring at a height of 
2 stellar radii above the surface.  The outer gap model spectrum of
Romani (1996) has an index of about -1.3 in the RXTE range, but falls
somewhat above the RXTE spectral points and does not predict a low energy 
cutoff or turnover.  The model of Cheng \& Zhang (1999),
in which particles from an outer gap accelerator travel toward the neutron
star surface emitting curvature radiation, produces a non-thermal hard
X-ray component through synchrotron radiation from the ensueing pair cascade.
However, the predicted spectrum, having index -1.9, is much softer than our 
measured RXTE spectra.  

The fact that the RXTE spectrum joins smoothly to the OSSE spectrum
and falls significantly below the ROSAT spectrum provides strong
evidence for a thermal rather than non-thermal origin for the ROSAT
emission.  The RXTE and ROSAT pulse profiles (see Figure~\ref{fig:velalcs}) are also very 
different, indicating separate origins for the emission.  A thermal 
component would most likely come from the hot
neutron star surface or polar cap.  This would be at odds with the outer 
gap model of Romani \& Yadigaroglu (1995), who interpret the ROSAT pulse as
non-thermal emission originating in acceleration regions in the outer
magnetosphere.

\begin{figure}
\plotone{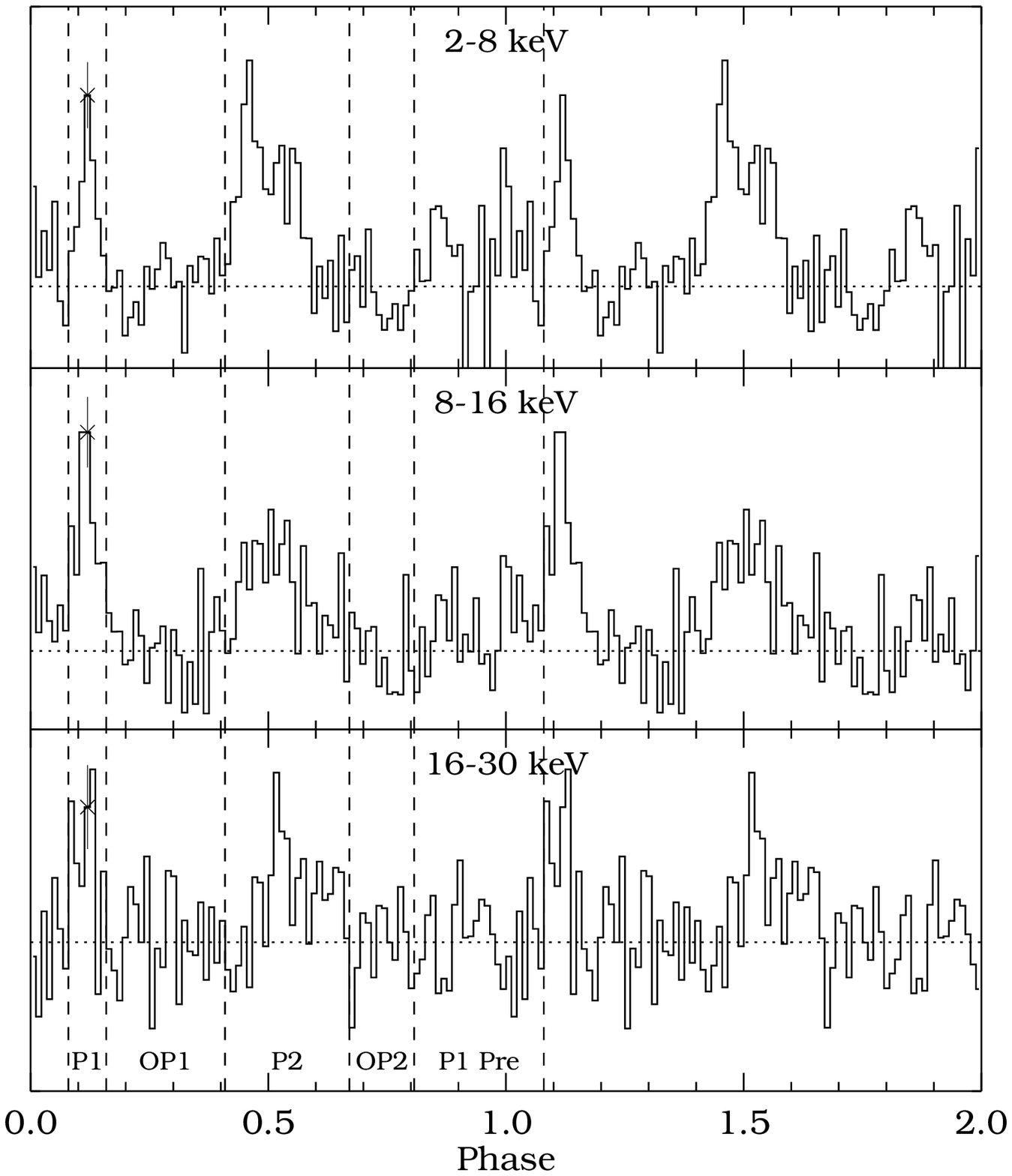}
\caption{Vela pulsar phase histograms in several broad energy bands,
averaged over the entire observation.  The vertical scale is arbitrary, but
the dotted line represents the average of the ``off-pulse'' regions.  Each
phase histogram displays a single
representative error bar for that band.  The phase region labels P1Pre, P1 and
P2 refer to Peak 1 Precursor, Peak 1 and Peak 2 respectively.  The regions OP1
and OP2 are the two ``off pulse'' regions.} 
\label{fig:pcalcs}
\end{figure}

\begin{figure}
\plotone{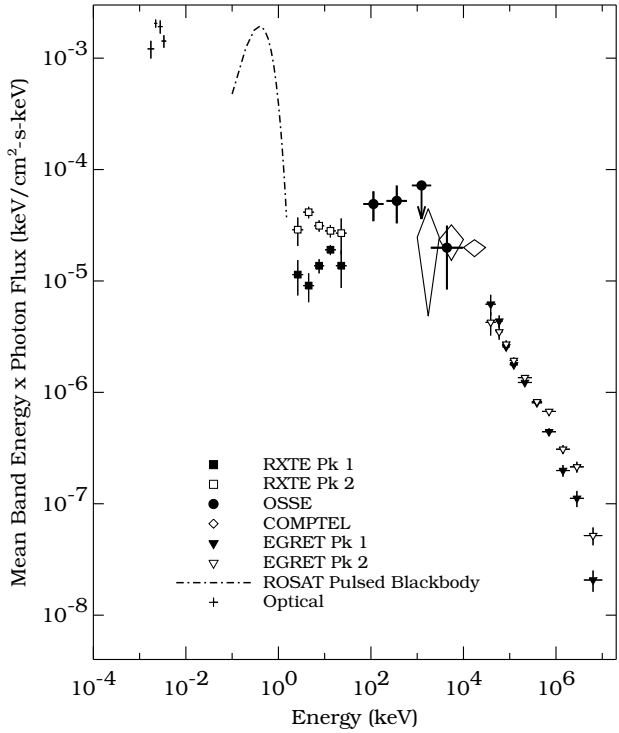}
\caption{Energy spectrum of pulsed high-energy and optical emission from the Vela Pulsar. 
In addition to the PCA spectra from this work, data shown include results from
optical (Nasuti et al. 1997), ROSAT (\"{O}gelman 1993), 
OSSE (Strickman et al. 1996), COMPTEL (Sch\"{o}nfelder et
al. 1994), and EGRET (Kanbach et al.
1994).  All results but optical are pulsed flux averaged over the entire light
curve.  The optical points are total emission from the point source.}   
\label{fig:velaspec}
\end{figure}

It is evident from Figure~\ref{fig:velaspec} that the averaged RXTE spectrum would 
extrapolate several orders of magnitude below the optical points.
However, the spectrum of Peak 2 is significantly softer than that of
Peak 1.  The spectrum of Peak 2, disregarding the lowest energy point
which has a large error bar, would extrapolate to within a factor of
two of the optical points.  It would then not connect to the OSSE
spectrum and would be a separate, low-energy conponent.  Although
the evidence for such a separate component in the Vela spectrum at
low energies is speculative at best from the spectral data alone, there
is possibly additional evidence to support such a picture from the
pulse profiles.  Figure~\ref{fig:velalcs} shows that the RXTE Peak 2 maximum seems to
line up with the optical peak 2.  The weaker secondary maximum in RXTE
Peak 2 is in phase with the OSSE and EGRET second peaks.  The phase
of Peak 2, as discussed in Section \ref{sec:anal} and shown in
Figure~\ref{fig:pcalcs}, 
appears to move higher with increasing energy. It is in phase alignment with the second 
optical pulse in the lowest energy band of 2 - 8 keV and in phase alignment 
with the second EGRET pulse in the highest energy band of 16-30 keV. 
The RXTE pulse may in fact be a blend of high and low energy 
components having 
different spectra, which could create the illusion of a peak moving with
energy.  While analysis of the present data using such a model
would not be conclusive, we shortly expect to obtain an additional 
300 ks of Vela observations with RXTE which should clarify the picture.

Finally, we point out that the RXTE Peak 1 precursor seems to be in phase 
with the radio peak (at phase 0 in Figure~\ref{fig:velalcs}).  The Peak 1 precursor has
the softest spectrum of any pulse profile component and has disappeared
into the background in the high-energy RXTE band (see Figure~\ref{fig:pcalcs}).
There is also a peak in the
optical profiles as well as a very weak and formally insignificant peak 
in the ROSAT profile in this phase region.  Correlated RXTE and radio observations, which we expect to complete in
the near future, will explore this connection.

\acknowledgments
We would like to thank the staff of the RXTE Guest Observer Facility for
invaluable aid in performing these analyses.  In particular, Arnold Rots was
extremely cooperative and helpful.  This work was supported in part by the
NASA/RXTE Guest Investigator program.

\newpage
\pagestyle{empty}
\clearpage

\clearpage 
\normalsize
\begin{table}  
\caption{Vela Pulsar Observation Summary}  \label{tbl:velaobs}
\centering 
\begin{tabular}{llr} 
StartTime (TJD) & Stop Time (TJD) & Exposure (ks)\\ \tableline
50460.563&50460.643&4.4\\
50461.564&50461.692&7.2\\
50465.429&50465.761&17.7\\
50465.865&50465.984&5.4\\
50466.525&50466.657&6.6\\
50468.379&50468.649&11.7\\
50469.371&50469.731&16.8\\
50470.439&50470.736&14.8\\
50471.590&50471.740&8.4
\end{tabular}
\end{table}

\normalsize
\begin{table}  
\caption{Vela Pulsar Ephemeris}  \label{tbl:velaephem}
\centering 
\begin{tabular}{ll} 
$\nu$ (s$^{-1}$)&$11.1961237576948$\\
$\dot{\nu}$ (s$^{-2}$)&$-1.56671\times10^{-11}$\\
$\ddot{\nu}$ (s$^{-3}$)&$3.27\times10^{-21}$\\
${\rm t_{0geo}}$ (MJD)&50440.000000226
\end{tabular}
\end{table}

\begin{table}
\caption{Peak and Off-pulse Phase Region Summary}
\label{tbl:phsregions}
\centering
\begin{tabular}{rrr}
Region & Start Phase & Phase Width\\
\tableline
 & &\\
Peak 1 Precursor & 0.807 & 0.272\\
Peak 1 & 0.079 & 0.080\\
Peak 2 & 0.409 & 0.262\\
Off-Pulse 1 & 0.159 & 0.250\\
Off-Pulse 2 & 0.671 & 0.136\\
\end{tabular}
\raggedright
\vspace{.2in}
\end{table}

\clearpage
\begin{table}
\caption{Vela Pulsar Maximum Likelihood Ratio (MLR) Test Summary} \label{tbl:velamlr}
\centering
\begin{tabular}{rrrr}
Energy Band & Peak 1 MLR $N_{\sigma}$& Peak 2 MLR $N_{\sigma}$&
Peak 1 Precursor MLR $N_{\sigma}$\\
\tableline
 & & & \\
2 -- 8 keV & 6.7 & 
8.9 & $ 3.5 $\\
8 -- 16 keV & 9.4 & 
6.4 & 2.4\\
16 -- 32 keV & 5.0 & 
3.7 & 0.2\\
\end{tabular}
\raggedright
\vspace{.2in}
\end{table}

\end{document}